\documentclass{elsart}

\usepackage{graphicx}
\usepackage{amssymb}

\begin{document}
\begin{frontmatter}

\journal{SCES'2001: Version 1}

\title{Tunneling current characteristics in bilayer quantum Hall systems}

%

\author[ut,iu]{Yogesh N. Joglekar}
\author[ut,iu]{Allan H. MacDonald}

 
\address[ut]{Department of Physics, University of Texas at Austin, Austin, TX 78705 }
\address[iu]{Department of Physics, Indiana University, Bloomington, IN 47405 }

%
%

\thanks[nsf]{This work was supported by the NSF under grant DMR0115947 and by
Robert A. Welch Foundation.}

%



\begin{abstract}
Weakly disordered bilayer quantum Hall systems at filling factor $\nu=1$ show spontaneous 
interlayer phase coherence if the layers are sufficiently close together. We study the 
collective modes in the system, the current-voltage characteristics and their evolution 
with an in-plane magnetic field in the phase-coherent regime. 
\end{abstract}



\end{frontmatter}


\noindent\emph{Introduction:}
Weakly disordered bilayer quantum Hall system has been a topic of intense research over 
the last decade~\cite{qhereviews,haf,apb,sqm}. This system consists of two 2D electron 
layers separated by a distance $d$ comparable to the distance between electrons within one 
layer. In a strong magnetic field perpendicular to the layers, because of the quenched kinetic 
energy, physical properties of the system are largely determined by the Coulomb 
interactions. For sufficiently small layer separation, bilayer quantum Hall system at 
\emph{total} filling factor $\nu=1$ undergoes a phase transition from a compressible state 
to an incompressible quantum Hall state with spontaneous phase-coherence~\cite{apb,sqm}. 
This phase-coherent state can be described as an easy-plane ferromagnet in pseudospin 
language where we associate a pseudospin with the layer index. Analogous to the spin-waves 
in a ferromagnet, the collective modes in this system are pseudospin-waves which transfer 
charge from one layer to the other. The properties of these collective modes are directly 
probed by interlayer tunneling experiments~\cite{spe}. In this paper we present an 
approximate but fully microscopic theory of interlayer current characteristics which takes 
into account the contribution of these collective modes.

\noindent\emph{Formalism:}
We consider a bilayer system with layer separation $d$ in a perpendicular magnetic field 
$B$ and an in-plane field $B_{||}$ confined to the region between the layers. In the 
strong-field limit, the pseudospin and the intra-Landau level index are the only dynamical 
degrees of freedom. The Hamiltonian for such a system with tunneling amplitude $\Delta_t$ 
consists of a one-body tunneling term $\hat{H}_t$, a one-body disorder potential 
$\hat{V}_{dis}$ and the two-body Coulomb interactions $\hat{V}_c$~\cite{ynj}. Due to the 
in-plane field, the tunneling matrix elements between the layers acquire Aharonov-Bohm 
phases which vary with wavevector $Q=dB_{||}/Bl^2$ where $l$ is the magnetic 
length~\cite{kyang}. Since the disorder arises from the impurities far away from the 
layers, we assume that it does not scatter electrons from one layer to the other and 
restrict ourselves to potentials with zero interlayer correlations~\cite{ynj}. 

\begin{figure}
\centering
\includegraphics[height=9cm,width=7.2cm]{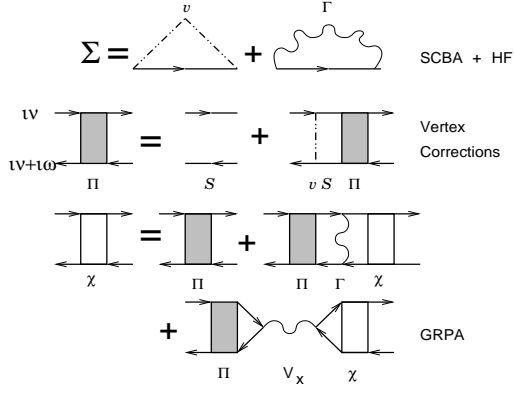}
\vspace{-1.5cm}
\caption{Diagrammatic summary of SCBA and GRPA. It is necessary to include vertex 
corrections along with the disorder broadening of quasiparticle bands. The competing 
Hartree and exchange fluctuations are captured by the direct ($V_x$) and exchange 
($\Gamma$) ladder diagrams.} 
\label{fig: feynman}
\end{figure}

The tunneling Hamiltonian $\hat{H}_t$ is the only term in the microscopic Hamiltonian which
 changes the layer-index of electrons. When the tunneling is zero, the charges in 
individual layers are conserved \emph{separately} and there is no interlayer current. 
Using Fermi's golden rule to estimate the rate of change of electron-number in each layer, 
we get
\begin{equation}
\label{eq: formal1}
I(V,B_{||})=\frac{e\Delta_t^2A}{\hbar}\mbox{ Im }\chi^{TB}(Q,eV)|_{\Delta_t=0}
\end{equation}
where $\chi^{TB}(\tau)=\langle T c^{\dagger}_T(\tau)c_B(\tau)c^{\dagger}_Bc_T\rangle$ is 
the thermal response function, $T(B)$ stands for the top(bottom) layer index and $A$ is the
 area of the sample. Eq.(\ref{eq: formal1}) relates the current, an observable, to the 
two-particle response function $\chi^{TB}$. We calculate this response function using the 
self-consistent Born approximation (SCBA) with vertex corrections and the generalized 
random phase approximation (GRPA) (Fig.~\ref{fig: feynman}). Due to the properties of 
Landau level wavefunctions, it is possible to evaluate the diagrams in 
Fig.~\ref{fig: feynman} analytically~\cite{ynj}. In the clean limit, $\mbox{Im }\chi$ is 
saturated by $\delta$-functions at $eV=\pm E_{sw}^Q$ where $E_{sw}^Q$ is the 
pseudospin-wave energy at wavevector $Q$~\cite{haf,apb}. In the presence of disorder, 
Eq.(\ref{eq: formal1}) predicts that the peak in the $I$-$V$ characteristics will be at 
voltages $eV=\pm E_{sw}^Q$. Thus we find that the collective-mode dispersion can be mapped 
out by varying the wavevector $Q$ or equivalently the in-plane field $B_{||}$~\cite{all}.

\noindent\emph{Results:} 
In absence of disorder, the mean-field approximation enhances the symmetric-antisymmetric 
splitting $\Delta_{SAS}=\Delta_t+\Delta_{sb}$~\cite{haf,apb}. The enhancement $\Delta_{sb}$
 survives in the limit $\Delta_t\rightarrow 0$ giving rise to spontaneous phase-coherence. 
We characterize the phase-coherent state by a dimensionless order parameter 
$M_0=n_S-n_{AS}$ where $n_\sigma$ is the integrated $\sigma$-state spectral-weight below 
the Fermi energy ($\sigma=S,AS$). We find that disorder-averaging broadens the sharp bands 
and leads to suppression of the order parameter from its clean-limit value, $M_0=1$. 

\begin{figure}
\centering
\includegraphics[height=7cm,width=7cm]{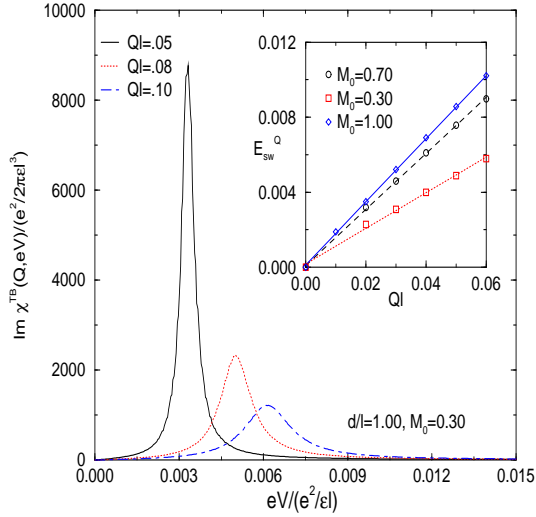}
\caption{Typical $I$-$V$ characteristics for various in-plane fields. The disorder strength
 is characterized by reduction of the order parameter from its clean-limit value, $M_0=1$. 
The position and the width of the peak are related to the pseudospin-wave energy 
$E_{sw}^Q\propto Q$, and and its damping, $\Gamma^Q\propto Q^2$, respectively. The inset 
shows typical pseudospin-wave dispersions for various disorder strengths. This linearly 
dispersing Goldstone mode has been recently observed by Spielman {\it et al.}~\cite{spe}.} 
\label{fig: stiffness}
\end{figure}

We obtain the $Q$-dependence of the pseudospin-wave energy $E_{sw}^Q$ and its damping 
$\Gamma^Q$ because of disorder from the real and imaginary parts of susceptibility $\chi$ 
respectively. The inset in Fig~\ref{fig: stiffness} shows typical dispersions for various 
disorder strengths. We find that disorder suppresses the pseudospin-wave velocity from its 
clean-limit value. We also find that at long wavelengths the damping term grows 
quadratically with the wavevector, $\Gamma^{Q}\propto Q^2$. In particular, \emph{even in 
the presence of disorder, the Goldstone mode at $Q=0$ is undamped}. These results agree 
with those obtained from effective-field-theory models, and the fact that this 
approximation captures these salient features is \emph{a} rationale behind choosing this 
particular set of diagrams. Furthermore, with this approximation we can obtain the energy 
and the width of the collective-mode at \emph{any} wavevector. The evolution of a typical 
$I$-$V$ curve with increasing in-plane field is shown in Fig.~\ref{fig: stiffness}. As 
expected from Eq.(~\ref{eq: formal1}) the tunneling current has a peak at the 
pseudospin-wave energy $E_{sw}^Q$. The finite width of the peak is because of the nonzero 
damping $\Gamma^Q$. We also find that for a fixed wavevector, the peak broadens with 
increasing disorder.

\noindent\emph{Summary:}
We have presented a microscopic theory of tunneling current in a disordered 
$\nu=1$ bilayer, which includes the effect of collective modes. We find that the current 
has a maximum at the collective-mode energy $E_{sw}^Q$. This treatment is valid only for 
nonzero wavevectors. We find that the collective mode is \emph{undamped} at zero wavevector
 leading to a breakdown of the perturbation theory in $\Delta_t$. Therefore, in such a 
case, a non-perturbative calculation in the tunneling amplitude is necessary \cite{ynj}.



\end{document}